\documentclass[prd,showpacs,floatfix,twocolumn,,amsmath,amssymb,floatfix]{revtex4}
\usepackage{graphicx,color,dcolumn,booktabs,bm}
\usepackage{longtable,lscape}
\usepackage{txfonts}
\usepackage{overpic}
\usepackage{amssymb}
\usepackage{indentfirst}
\usepackage{epsfig}
\usepackage{epstopdf}   
\usepackage{slashed}  
\usepackage{cases}
\usepackage{color}
\usepackage{multirow}
\usepackage[section]{placeins}
\usepackage{graphicx,color,dcolumn,booktabs,bm}

\graphicspath{{Figures/}} %

\usepackage[colorlinks, citecolor=blue,anchorcolor=red,menucolor=red, linkcolor=red,filecolor=red,runcolor=red,urlcolor=blue,frenchlinks=red]{hyperref}

\usepackage[none]{hyphenat} 

\usepackage{color,xcolor,fancybox,epsf,rotating,colordvi}

\newcommand{\spl}{{^3P_0}}
\newcommand{\kzst}{{K_0^*}}

\newcommand{\bdsp}{{\boldsymbol{p}}}

\begin{document}

\title{The possible assignments of the scalar $K_0^*(1950)$ and $K_0^*(2130)$ within the $^3P_0$ model }

\author{Tian-Ge Li}
\affiliation{Joint Research Center for Theoretical Physics, School of Physics and Electronics, Henan University, Kaifeng 475004, China}

\author{Zhuo Gao}
\affiliation{Joint Research Center for Theoretical Physics, School of Physics and Electronics, Henan University, Kaifeng 475004, China}

\author{Guan-Ying Wang}\email{wangguanying@henu.edu.cn}
\affiliation{Joint Research Center for Theoretical Physics, School of Physics and Electronics, Henan University, Kaifeng 475004, China}

\author{De-Min Li}\email{lidm@zzu.edu.cn}
\affiliation{School of Physics and Microelectronics, Zhengzhou University, Zhengzhou, Henan 450001, China}

\author{En Wang}\email{wangen@zzu.edu.cn}
\affiliation{School of Physics and Microelectronics, Zhengzhou University, Zhengzhou, Henan 450001, China}

\author{Jingya Zhu}\email{zhujy@henu.edu.cn}
\affiliation{Joint Research Center for Theoretical Physics, School of Physics and Electronics, Henan University, Kaifeng 475004, China }

\date{\today}

\begin{abstract}
We have evaluated the strong decays of the $K_0^*(1950)$ and $K_0^*(2130)$ within the $^3P_0$ model,  by employing the meson wave functions from the relativized quark model. By comparing with the experimental measurements, the $K_0^*(2130)$ could be assigned as $K_0^*(3^3P_0)$, while the $K_0^*(1950)$ seems like an exotic state,  because its width can not be reasonably reproduced within the $^3P_0$ model.
We also predict that the $K_0^*(2^3P_0)$ state has a mass of about $1811$~MeV and a width of about $656$~MeV, while the $K_0^*(4^3P_0)$ state has a mass of about $2404$~MeV and a width of about $180$~MeV.
\end{abstract}
\pacs{ }
\maketitle

\section{Introduction}{\label{Introduction}}
According to the theory of quantum chromodynamics (QCD), in addition to the conventional $q\bar{q}$ mesons, the so-called exotic states  are  also permitted, such as tetraquarks, molecules, glueballs, and hybrids~\cite{Richard:2016eis,Jaffe:2004ph,Amsler:2004ps}. Although many exotic states have been observed experimentally, such as $X(3872)$, $Z_c(3900)$, $Z_c(4025)$, $P_c$, $T_{cc}$, it is still difficult to distinguish between the exotic states with conventional quantum numbers and the ordinary $q\bar{q}$ mesons.

One  puzzle in hadron spectra is the scalar mesons, since there are too many states to be accommodated within the quark model without difficulty \cite{tHooft:2008rus}. For example,  $\kzst(700)$ state (also known as $\kappa$), together with its multiple partners $a_0(980)$, $f_0(500)(\sigma)$, and $f_0(980)$, does not fit well into the predictions of the quark model, since the observed mass
ordering of these lowest scalar states is $m_\sigma < m_\kappa < m_{a_0,f_0}$ \cite{Zyla:2020zbs},
while in the conventional quark model, by a naive counting of the quark mass,
the mass ordering of the scalar $q\bar{q}$ nonet should be $m_\sigma \sim m_{a_0} < m_\kappa < m_{f_0}$. These scalar states below 1~GeV are generally believed not to be $q\bar{q}$ states \cite{Close:2002zu, Maiani:2004uc, Amsler:2004ps, Pelaez:2003dy, Jaffe:2004ph, Eichmann:2015cra}.

Within the naive quark model,  it is natural to assume that the $a_0(1450)$, $\kzst(1430)$, $f_0(1710)$, and $f_0(1370)$ are the $1\spl$ members of the SU(3) flavor nonet \cite{Zyla:2020zbs}.
The isovector scalar mesons $a_0(2020)/a_0(1950)$ are suggested to be the good candidates of the $a_0(3\spl)$ in our previous work \cite{Wang:2017pxm}. However, the assignments of the excited  scalar $\kzst$ states are still unclear. Up to now, above the $\kzst(1430)$ mass, only two scalar states $\kzst(1950)$ and $\kzst(2130)$ are reported. Their masses and widths are listed in Table~\ref{tab:pdg}
.

\begin{table}[tpbh]
	\centering
	\caption{The masses and widths of the excited scalar $K_0^\ast$ states (in MeV).}
	\label{tab:pdg}
	\begin{tabular}{c c c c}
		\hline\hline
		state          &mass                 & width  & ref \\
		\hline
		$\kzst(1950)$   &1945$\pm$10$\pm$20   &201$\pm$34$\pm$79   & \cite{Zyla:2020zbs}  \\
        $\kzst(2130)$   &2128$\pm$31$\pm$9    &95$\pm$42$\pm$76   & \cite{Lees:2021cww} \\

		\hline\hline
	\end{tabular}
\end{table}

Since $\kzst(1950)$ was first reported in the $K\pi$ invariant mass distribution of the $K^-p\to K^-\pi^+ n$ reaction by LASS in 1988 \cite{Aston:1987ir}, it is difficult to interpret its properties within the quark model. The $\kzst(1950)$ mass is close to  the $\kzst(2\spl)$ mass of about 1890~MeV predicted by the Godfrey-Isgur (GI) quark model~\cite{Godfrey:1985xj}. However, it is expected that the  $\kzst(2\spl)$ with a mass of 1850~MeV has a width of  about 450~MeV within the $\spl$ decay model~\cite{Barnes:2002mu}, larger than the $\kzst(1950)$ width.
In addition, Ref.~\cite{Pang:2017dlw} recently analyzed the kaon family within the modified GI model involving the color screening effect, and predicted the mass and width of the $\kzst(2\spl)$ to be $M=1829$~MeV and $\Gamma=1000$~MeV, respectively, both of which disfavor the assignment of $\kzst(1950)$ as the candidate the $\kzst(2\spl)$ state.

The $\kzst(2130)$ was recently observed in the ${\eta}_c$ decays by the {\it BABAR} Collaboration~\cite{Lees:2021cww}, and its mass is close to the $\kzst(3\spl)$ mass of 2176~MeV  predicted by the modified GI model~\cite{Pang:2017dlw}.
In addition, we have estimated that the $n\bar{n}(3^3P_0)$ mass is about $1.9\sim2.0$~GeV~\cite{Wang:2017pxm}, thus one can naturally expect  the  $\kzst(3\spl)$ mass should be about $100\sim200$ MeV larger than the $a_0(3\spl)$ mass.  Based on its mass information, the $\kzst(2130)$ seems a good candidate of the $\kzst(3\spl)$.

The mass information alone is insufficient to identify the $\kzst(2130)$ as the $\kzst(3\spl)$ state.
We shall discuss the possibility of the $\kzst(2130)$ as the $\kzst(3\spl)$ state by studying its strong decay properties.

In this work, we will investigate the possible assignment of  $\kzst(2130)$ by analyzing the strong decay behaviors within the $\spl$ decay model. For completeness, we also check the possibility of the $\kzst(1950)$ as the ordinary scalar mesons, since it is natural and necessary to exhaust the possible $q\bar{q}$ descriptions of a newly observed state before restoring to the more exotic assignments.
This paper is organized as follows. In Sec.~\ref{sec:formalisms}, we introduce the $\spl$ strong decay model used in our calculations, and the results and discussions are given in Sec.~\ref{sec:result}. Finally, a summary is given in Sec.~\ref{sec:summary}.

\section{Model and Parameters}
\label{sec:formalisms}

\begin{figure}[htpb]
\includegraphics[scale=0.5]{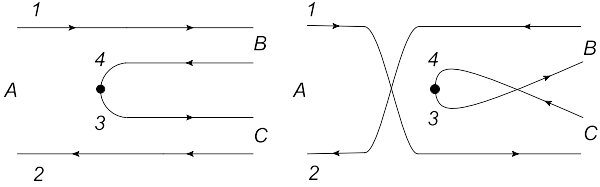}
\vspace{0.0cm}\caption{Two-body-decay diagrams of $A\to BC$ according to the $\spl$ model, where a pair of quark-antiquark are created to form the final two mesons.
In the left diagram, Meson $B$ is formed by quark in Meson $A$ combined with the created antiquark, and Meson $C$ is formed by antiquark in Meson $A$ combined with the created quark.
In the right diagram, Meson $B$ is formed by antiquark in Meson $A$ combined with the created quark, and Meson $C$ is formed by quark in Meson $A$ combined with the created antiquark.
}\label{fig:feynman}
\end{figure}

The $\spl$ model has been widely used to study the Okubo-Zweig-Iizuka (OZI)-allowed open flavor two-body strong decays,
it was originally introduced by Micu~\cite{Micu:1968mk} and further developed by Le Yaouanc $et$ $al.$~\cite{LeYaouanc:1972vsx,LeYaouanc:1973ldf,LeYaouanc:alo}.
In the $\spl$ model, the meson strong decay takes place by producing a quark-antiquark pair with vacuum quantum number $J^{PC}=0^{++}$. The newly produced quark-antiquark pair, together with the $q\bar{q}$ within the initial meson, regroups into two outgoing mesons in two possible quark rearrangement ways, as shown in Fig.~\ref{fig:feynman}. The $^3P_0$ model has been widely applied to study strong decays of hadrons with considerable success ~\cite{Roberts:1992js,Blundell:1996as,Barnes:1996ff,Barnes:2002mu,
Close:2005se,Barnes:2005pb,Zhang:2006yj,
Li:2009rka,Li:2010vx,Lu:2014zua,Pan:2016bac,Lu:2016bbk, Wang:2017pxm,Li:2021qgz,Hao:2019fjg,Xue:2018jvi}.

Following the conventions in Refs.~\cite{Roberts:1992js,Blundell:1996as}, the transition operator $T$ of the decay  $A\rightarrow BC$ in the $\spl$ model is given by
\begin{eqnarray}
T&=&-3\gamma\sum_m\left< 1m1-m|00\right>\int
d^3\bdsp_3d^3\bdsp_4\delta^3(\bdsp_3+\bdsp_4)\nonumber\\
&&{\cal{Y}}^m_1\left(\frac{\bdsp_3-\bdsp_4}{2}\right
)\chi^{34}_{1-m}\phi^{34}_0\omega^{34}_0b^\dagger_3(\bdsp_3)d^\dagger_4(\bdsp_4),
\end{eqnarray}
where the $\gamma$ is a dimensionless parameter corresponding to the production strength of the quark-antiquark pair $q_3\bar{q}_4$ with quantum number $J^{PC}=0^{++}$. $\bdsp_3$ and  $\bdsp_4$ are the momenta of the created quark  $q_3$ and  antiquark $\bar{q}_4$, respectively. $\chi^{34}_{1,-m}$, $\phi^{34}_0$, and $\omega^{34}_0$ are the spin, flavor, and color wave functions of $q_3\bar{q}_4$ system, respectively. The solid harmonic polynomial  ${\cal{Y}}^m_1(\bdsp)\equiv|\bdsp|^1Y^m_1(\theta_p, \phi_p)$ reflects the momentum-space distribution of the $q_3\bar{q_4}$.

The $S$ matrix of the process $A\rightarrow BC$ is defined by
\begin{eqnarray}
\langle BC|S|A\rangle=I-2\pi i\delta(E_A-E_B-E_C)\langle BC|T|A\rangle,
\end{eqnarray}
where $|A\rangle$ ($|B\rangle$, $|C\rangle$) are the wave functions of the mock mesons defined by Ref.~\cite{Hayne:1981zy}.

The transition matrix element $\langle BC|T|A\rangle$ can be written as
\begin{eqnarray}
\langle BC|T|A\rangle=\delta^3(\bdsp_A-\bdsp_B-\bdsp_C){\cal{M}}^{M_{J_A}M_{J_B}M_{J_C}}(\bdsp),
\end{eqnarray}
where ${\cal{M}}^{M_{J_A}M_{J_B}M_{J_C}}
(\bdsp)$ is the helicity amplitude.

The partial wave amplitude ${\cal{M}}^{LS}(\bdsp)$ can be given by~\cite{Jacob:1959at},
\begin{eqnarray}
{\cal{M}}^{LS}(\bdsp)&=&
\sum_{\renewcommand{\arraystretch}{.5}\begin{array}[t]{l}
\scriptstyle M_{J_B},M_{J_C},\\\scriptstyle M_S,M_L
\end{array}}\renewcommand{\arraystretch}{1}\!\!
\langle LM_LSM_S|J_AM_{J_A}\rangle \nonumber\\
&&\langle
J_BM_{J_B}J_CM_{J_C}|SM_S\rangle\nonumber\\
&&\times\int
d\Omega\,\mbox{}Y^\ast_{LM_L}{\cal{M}}^{M_{J_A}M_{J_B}M_{J_C}}
(\bdsp). \label{pwave}
\end{eqnarray}

Various $\spl$ models exist in literature and typically differ in the choices of the pair-production vertex, the phase space conventions, and the meson wave functions employed. In this work, we restrict to the simplest vertex as introduced originally by Micu \cite{Micu:1968mk} which assumes a spatially constant pair-production strength $\gamma$, adopt the relativistic phase space, and employ the relativized quark model (RQM) wave functions \cite{Godfrey:1985xj}.

With the relativistic phase space, the decay width
$\Gamma(A\rightarrow BC)$ can be expressed in terms of the partial wave amplitude
\begin{eqnarray}
\Gamma(A\rightarrow BC)= \frac{\pi
|\bdsp|}{4M^2_A}\sum_{LS}|{\cal{M}}^{LS}(\bdsp)|^2, \label{width1}
\end{eqnarray}
where $|\bdsp|=\sqrt{[M^2_A-(M_B+M_C)^2][M^2_A-(M_B-M_C)^2]}/(2M_A)$,
and $M_A$, $M_B$, and $M_C$ are the masses of the mesons $A$, $B$,
and $C$, respectively.

\section{RESULTS}
\label{sec:result}

In our calculations, the parameters involve the $q\bar{q}$ pair production strength $\gamma$, and the ones in relativized quark model, as used in the work of Godfrey and Isgur~\cite{Godfrey:1985xj}.
The flavor wave functions for the mesons are adopted by following the conventions of Refs.~\cite{Barnes:2002mu, Godfrey:1985xj} except for (1) $f_1(1285)=-0.28 n\bar{n}+0.96 s\bar{s}$ and $f_1(1420)=-0.96 n\bar{n}-0.28 s\bar{s}$ as Ref.~\cite{Li:2000dy}, (2) $\eta(1295)=(n\bar{n}-s\bar{s})/\sqrt{2}$ and $\eta(1475)=(n\bar{n}+s\bar{s})/\sqrt{2}$ as Ref.~\cite{Yu:2011ta},  where $n\bar{n}=(u\bar{u}+d\bar{d})/\sqrt{2}$.  The masses of the final mesons are taken from the Review of Particle Physics~\cite{Zyla:2020zbs}.

We take $\gamma = 0.52$ by fitting to the total width of $\kzst(1430)$ as the $1\spl$ state. The decay widths of $\kzst(1430)$ as the $\kzst(1\spl)$ state are listed in Table~\ref{tab:13p0}. According to our results, the dominant decay mode of $\kzst(1430)$ is $K\pi$, which is consistent with the experimental data \cite{Zyla:2020zbs}.

\begin{table}[h]
\begin{center}
	\caption{Decay widths of $\kzst(1430)$ as $1\spl$ state (in MeV).}
	\label{tab:13p0}
	\begin{tabular}{c| c c}
		\hline\hline
		 Channel                      & Mode                  &$\Gamma_i(1\spl)$    \\
		\hline
		$0^+\rightarrow 0^-+0^- $     & $K\pi$                &262.53           \\
		                              & $K\eta$               &9.42             \\	
		$0^+\rightarrow 0^-+1^+ $     & $\pi K_{1B}$          &$<0.1$             \\
		\hline	
		                              & Total width           &271.95         \\
		  \hline
		                              & Experiment            & $270\pm80$        \\
		\hline\hline
	\end{tabular}
\end{center}
\end{table}

With the parameter $\gamma = 0.52$, we have calculated the decay widths of $\kzst(1950)$ as the $\kzst(2\spl)$ and $\kzst(3\spl)$ states, respectively, which are listed in Table~\ref{tab:RQM}. The total widths of the $\kzst(2\spl)$ and $\kzst(3\spl)$ states with a mass of 1945~MeV are expected to be  $1308$~MeV and $74$~MeV, respectively. By comparing with the $K^\ast_0(1950)$ width $\Gamma=201\pm34\pm79$ MeV~\cite{Zyla:2020zbs}, it is hard to assign the $\kzst(1950)$ as the ordinary scalar $q\bar{q}$ meson.

\begin{table}[h]
\begin{center}
	\caption{Decay widths of $\kzst(1950)$ as $2\spl$ and $3\spl$ states (in MeV). The initial state mass is $1945$ MeV.}
	\label{tab:RQM}
	\begin{tabular}{c |c |c c }
		\hline\hline
	Channel                      & Mode                  &$\Gamma_i(2^3P_0)$     & $\Gamma_i(3^3P_0)$   \\
	\hline
	$0^+\rightarrow 0^-+0^- $     & $\pi K$                &     78.26             &        1.64             \\
	& $K\eta$                                             &     1.74             &       $<0.01$               \\
	& $K\eta'$                                          &     11.39             &        1.37               \\
	& $\pi(1300) K$                                      &   170.82            &         6.79               \\
	& $K\eta(1295)$                                    &     147.80            &         4.15             \\
	& $\pi K(1460)$                                     &    91.15             &         0.34            \\
	$0^+\rightarrow 0^-+1^+ $     & $Ka_1 (1260)$          &    57.37            &      0.03              \\
	& $Kb_1  (1235)$                                    &    106.81             &       10.46             \\
	& $h_1 (1415)K$                                     &    11.70            &       2.77             \\
	& $h_1 (1170)K$                                     &     5.87             &       $<0.01$           \\
	& $Kf_1 (1420)$                                     &    2.05             &        0.02           \\
	& $Kf_1 (1285)$                                    &    4.59             &        $<0.01$              \\
	& $\pi K_{1A}$                                     &    132.10             &         1.13                  \\
	& $\pi  K_{1B}$                                     &    25.27              &         2.08                \\
	& $ \eta  K_{1B}$                                   &     54.80            &          6.70               \\
	$ 0^+\rightarrow1^-+1^- $     & $K^*(892)\rho $   &    204.87             &        18.60               \\
	& $K^*(892)\phi(1020)$                              &   9.17             &       0.04                 \\
	& $K^*(892)\omega$                                  &   192.59             &       17.84                \\
	$ 0^+\rightarrow0^-+2^- $    & $\pi K_2(1770)$   &  0.02                 &   0.03          \\
	\hline
	& Total width                                       &   1308.38            &         74.01               \\
	\hline
	& Experiment                                      & \multicolumn{2}{c}{201$\pm$34$\pm$79}  \\
	\hline\hline
	\end{tabular}
\end{center}
\end{table}

\begin{figure}[htpb]
\includegraphics[scale=0.7]{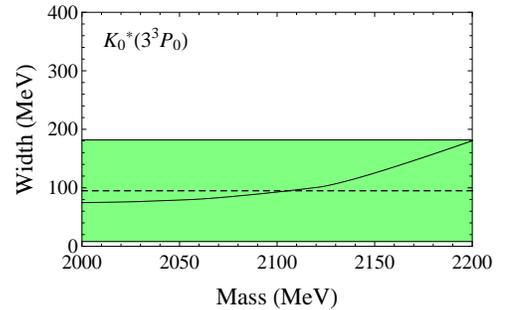}
\vspace{0.0cm}
\caption{The dependence of the total width of $\kzst(3\spl)$ on the initial state mass. The experimental total width of the $\kzst(2130)$ is denoted by the dashed line with a green band.}
\label{fig:33p0}
\end{figure}

\begin{figure}[htpb]
	\includegraphics[scale=0.7]{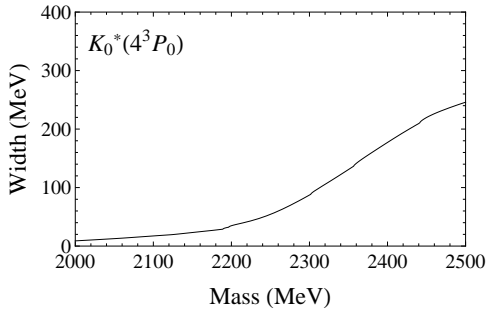}
	\vspace{0.0cm}\caption{The dependence of the total width of $\kzst(4\spl)$ on the initial state mass.}
	\label{fig:43P0}
\end{figure}

Then we will discuss the possible assignment of the $\kzst(2130)$ state. We have shown the decay widths of $\kzst(2130)$ as the $\kzst(3\spl)$ and $\kzst(4\spl)$ states in Table~\ref{tab:2130}, and total width is expected to be about $105$ MeV and $20$ MeV, respectively for the cases of the $\kzst(3\spl)$ and $\kzst(4\spl)$.
The dependences of the total widths of $\kzst(3\spl)$ and $\kzst(4\spl)$ on the initial state mass are shown in Fig.~\ref{fig:33p0} and Fig.~\ref{fig:43P0}, respectively. Within the uncertainties of the $K^*_0(2130)$ mass, the total width of $\kzst(3\spl)$ is in agreement with the experimental data $\Gamma= 95\pm42\pm76$ MeV~\cite{Zyla:2020zbs},  which implies that the $K^\ast_0(2130)$ should be the good candidate of the $\kzst(3\spl)$ state.

 \begin{table}[h]
 	\begin{center}
 		\caption{Decay widths of $\kzst(2130)$ as $3\spl$ and $4\spl$ states (in MeV). The initial state mass is $2128$ MeV.}
 		\label{tab:2130}
 		\begin{tabular}{c| c| c c }
 			\hline\hline
 			Channel                      & Mode             & $\Gamma_i(3^3P_0)$    &$\Gamma_i(4^3P_0)$\\
 			\hline
 			$0^+\rightarrow 0^-+0^- $    & $\pi K$          &          8.58           &      0.08         \\
 			& $K\eta$                                       &        0.07         &      $<0.01$          \\
 			& $K\eta'$                                      &         0.21         &     0.29          \\
 			& $\pi(1300) K$                                 &         4.35           &        4.09          \\
 			& $K\eta(1295)$                                 &        7.38           &      3.55         \\
 			& $\pi K(1460)$                                 &        8.83          &     2.83          \\
 			& $K\eta(1475)$                                 &       $<0.01$         &       0.05          \\
 			& $K(1460)\eta$                                 &         0.87          &     $<0.01$          \\
 			$0^+\rightarrow 0^-+1^+ $     & $Ka_1 (1260)$   &         3.58        &      0.09         \\
 			& $Kb_1  (1235)$                                &         1.63          &       1.52          \\
 			& $h_1 (1415)K$                                 &         12.41          &     0.02         \\
 			& $h_1 (1170)K$                                 &         0.84          &     0.11         \\
 			& $Kf_1 (1420)$                                 &         0.04          &    0.10         \\
 			& $Kf_1 (1285)$                                 &        0.16         &      $<0.01$         \\
 			& $\pi K_{1A}$                                  &        2.01           &      0.19          \\
 			& $\pi  K_{1B}$                                 &         0.06       &       3.02          \\
 			& $ \eta  K_{1B}$                               &         2.67         &       0.32          \\
 			& $ \eta  K_{1A}$                               &       3.26         &       0.04        \\
 			$ 0^+\rightarrow1^-+1^- $     & $K^*(892)\rho $ &         14.27           &     0.61          \\
 			& $K^*(892)\phi(1020)$                          &          1.42         &       0.08        \\
 			& $K^*(892)\omega$                              &       14.97          &       0.41         \\
 			$ 0^+\rightarrow1^++1^- $     & $K^*(892)a_1 (1260) $  &       1.20          &       0.16         \\
 			& $K^*(892)b_1  (1235)$                                &        0.15           &      0.02         \\
 			& $K^*(892)h_1 (1170)$                                 &         2.33         &    0.26          \\
 			& $\rho K_{1B}$                                        &        2.56          &      0.11        \\
 			& $ \omega K_{1B}$                                     &        2.87          &      0.30          \\
 			$ 0^+\rightarrow0^-+2^- $     & $\pi K_2(1820) $       &      1.28          &     0.11        \\
 			& $\pi K_2(1770)$                                      &        7.22          &      1.69        \\
 			& $\eta_2(1645) K$                                     &     0.05         &      0.02        \\
 			\hline
 			& Total width                                          &       105.33        &      20.07         \\
 			\hline
 			& Experiment                  &\multicolumn{2}{c}{ 95$\pm$42$\pm$76 }  \\
 			\hline\hline
 			
 		\end{tabular}
 	\end{center}
 \end{table}

\begin{figure}[htpb]
\includegraphics[scale=0.3]{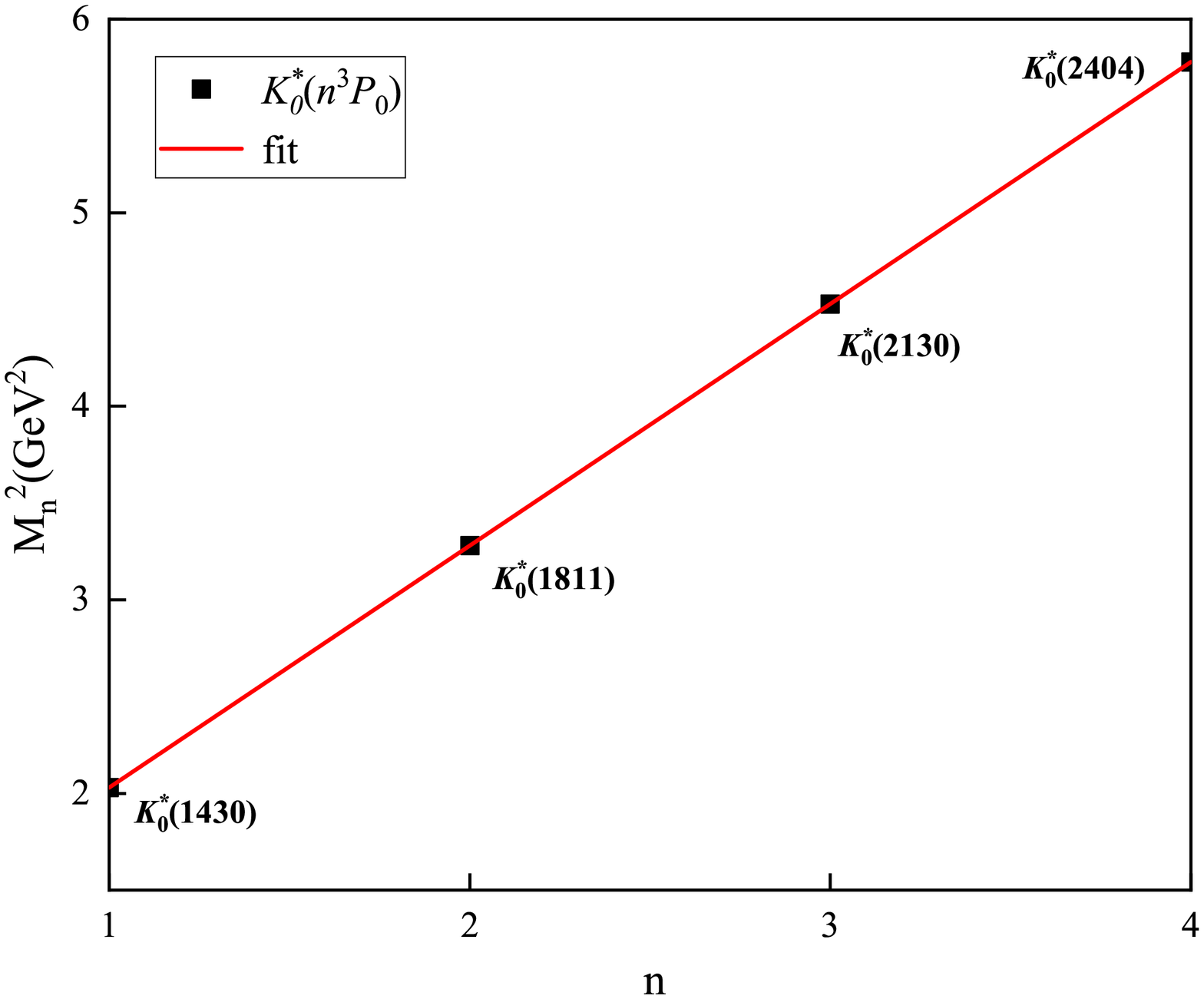}
\vspace{0.0cm}\caption{The Regge trajectory of the $\kzst(n\spl)$ ($n=1,2,3,4$) states, where $M_n$ is its mass, and the red line is the fitting result.}
\label{fig:regge}
\end{figure}

Phenomenologically, it is suggested that the light mesons could be grouped
into the following Regge trajectories~\cite{Anisovich:2000kxa,Pan:2016bac,Xue:2018jvi},
\begin{equation}
M_n^2=M_0^2 + (n-1) \mu^2,
\label{eq:regge}
\end{equation}
where $M_0$ is the lowest-lying meson mass, $n$ is the radial quantum number, and $\mu^2$ is the slope parameter of the corresponding trajectory.
In the presence of $\kzst(1430)$ and $\kzst(2130)$ being the $\kzst(1\spl)$ and $\kzst(3\spl)$ states, we can roughly estimate the $\kzst(2\spl)$ mass to be about 1811~MeV and $\kzst(4\spl)$ mass to be about 2404 MeV as shown in Fig.\ref{fig:regge}.

\begin{figure}[htpb]
	\includegraphics[scale=0.7]{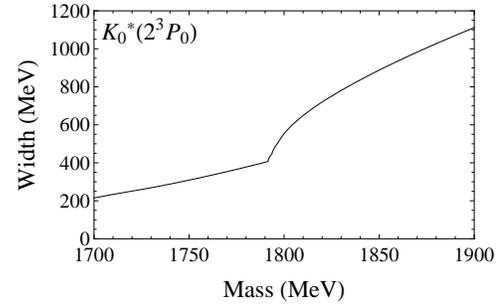}
	\vspace{0.0cm}\caption{The dependence of the total width of $\kzst(2\spl)$ on the initial state mass.}
	\label{fig:23P0}
\end{figure}

It should be pointed out that, in our previous work~\cite{Wang:2017pxm}, the mass scale for the $n\bar{n}(2^3P_0)$ nonets is expected to be $1700\sim 1800$~MeV. In addition, the mass of $\kzst(2\spl)$ state is predicted to be 1890 MeV by the GI model~\cite{Godfrey:1985xj} and to be 1829~MeV by the modified GI model~~\cite{Pang:2017dlw}.  The prediction of the $\kzst(2\spl)$ mass from the Regge trajectories is consistent with these previous predictions.
The strong decays of $\kzst(2\spl)$ with a mass of 1811~MeV are presented in Table~\ref{tab:23p0}.
The dependence of the total width of $\kzst(2\spl)$ on the initial state mass is shown in Fig.~\ref{fig:23P0}. When the initial state mass varies from 1700 MeV to 1900 MeV,  the total width of the $\kzst(2\spl)$ varies from about 215~MeV to 1111~MeV. The decay width varies greatly, since some decay modes are open gradually.
The total width of $\kzst(2\spl)$ is expected to be about 656~MeV. The dominant decay modes of $\kzst(2\spl)$ are $\pi(1300) K$, $K\eta(1295)$, and $\pi K(1460)$. The $\kzst(2\spl)$ is predicted to be a broad state, which could be the reason that the $\kzst(2\spl)$ candidate is not yet observed experimentally.

\begin{table}[h]
\begin{center}
	\caption{Decay widths of $\kzst(2\spl)$ (in MeV). The initial state mass is $1811$ MeV. }
	\label{tab:23p0}
		\begin{tabular}{c c  c}
			\hline\hline
			Channel                      & Mode                                    &$\Gamma_i(2\spl)$\\
			\hline
			$0^+\rightarrow 0^-+0^- $     & $\pi K$                                 &57.13  \\
				& $\pi(1300) K$                            &96.06\\
			& $K\eta$                                 &1.01 \\
			& $K\eta'$                                &0.77 \\
				& $K\eta(1295)$                           &100.95\\
			& $\pi K(1460)$                           &99.91  \\
			$0^+\rightarrow 0^-+1^+ $     & $h_1(1170)K$                            &4.42  \\
			  & $Ka_1 (1260)$                 &16.35\\
			  	& $Kf_1 (1285)$                      &0.42\\
			  	& $Kb_1  (1235)$                      &33.29\\
			& $\pi K_{1A}$                            &64.31 \\
			& $\pi  K_{1B}$                           &24.37\\
			$ 0^+\rightarrow1^-+1^- $     & $K^*(892)\rho $                         &81.42 \\
			& $K^*(892)\omega$                        &75.93 \\
		
			& Total width                             &656.34\\
			\hline\hline
		\end{tabular}
	\end{center}
\end{table}

 We show the partial decay widths and the total decay width of the $\kzst(4\spl)$ state with a mass of 2404~MeV in Table~\ref{tab:43P0}. The total width of $\kzst(4\spl)$ is expected to be about 180~MeV. The dominant decay modes of $\kzst(4\spl)$ include $K^*(1410)^+\omega$, $K^*(1410)^+\rho$, $K^*(892)\omega(1420)$. The dependence of the total width of the $\kzst(4\spl)$ state on the initial state mass is shown in Fig.~\ref{fig:43P0}. When the initial state mass varies from 2300 MeV to 2500 MeV, the total width of the $\kzst(4\spl)$ varies from about $88\sim 246$~MeV.

\begin{table}[h]
	\begin{center}
		\caption{Decay widths of $\kzst(4\spl)$ (in MeV). The initial state mass is $2404$ MeV. }
		\label{tab:43P0}
		\begin{tabular}{c |c|  c}
			\hline\hline
			Channel                      & Mode                &$\Gamma_i(4\spl)$\\
			\hline
			$0^+\rightarrow 0^-+0^- $    & $\pi K$                            &0.60\\
			& $K\eta$                                &        $<0.01$           \\
			& $K\eta'$                            &        0.28           \\
			& $\pi(1300) K$                            &         0.16           \\
			& $K\eta(1295)$                           &         0.07           \\
			& $\pi K(1460)$                         &        0.12           \\
			& $K\eta(1475)$                               &        $<0.01$          \\
			& $K(1460)\eta$                           &         0.31           \\
			$0^+\rightarrow 0^-+1^+ $     & $Ka_1 (1260)$                 &        0.70          \\
			& $Kb_1  (1235)$                      &         4.50           \\
			& $h_1 (1415)K$                         &        4.60           \\
			& $h_1 (1170)K$                      &        0.06           \\
			& $Kf_1 (1420)$                       &        0.21          \\
			& $Kf_1 (1285)$                       &         0.03           \\
			& $\pi K_{1A}$                       &         0.02          \\
			& $\pi  K_{1B}$                        &         1.69           \\
			& $ \eta  K_{1B}$                      &        2.74           \\
			& $ \eta  K_{1A}$                      &         0.98          \\
			& $ \eta'  K_{1B}$                      &       0.08           \\
			& $ \eta'  K_{1A}$                      &         0.01         \\
			& $Ka_1 (1640)$                 &       0.23          \\
			$0^+\rightarrow 0^++1^- $     & $K_0^*(1430)\rho$                 &       0.09          \\
			$ 0^+\rightarrow1^-+1^- $     & $K^*(892)\rho $                &         8.76          \\
			& $K^*(892)\rho(1450) $                &        6.34          \\
			& $K^*(892)\omega(1420)$                   &        19.83          \\
			& $K^*(892)\phi(1020)$                   &         0.02          \\
			& $K^*(892)\omega$                        &         8.15          \\
			& $K^*(1410)\rho $                &         32.46          \\
			
			& $K^*(1410)\omega$                        &        36.79         \\
			$ 0^+\rightarrow1^++1^- $     & $K^*(892)a_1 (1260) $                 &         2.50           \\
			& $K^*(892)b_1  (1235)$                     &         0.03           \\
			& $K^*(892)h_1 (1170)$                        &        0.13           \\
			& $K^*(892)h_1 (1415)$                        &       $<0.01$         \\
			& $K^*(892)f_1 (1420)$                        &        0.69           \\
			& $K^*(892)f_1 (1285)$                        &        0.34           \\
			
			& $\rho K_{1A}$                                &        $<0.01$           \\
			& $ \omega K_{1A}$                            &        $<0.01$        \\
			& $\phi K_{1B}$                                &         0.25           \\
			& $\rho K_{1B}$                                &         1.29           \\
			& $ \omega K_{1B}$                            &         0.22           \\
			$ 0^+\rightarrow0^-+2^- $     & $\pi K_2(1820) $                     &         0.17           \\
			& $\pi K_2(1770)$                           &         2.03           \\
			& $K\pi_2(1670)$                           &         7.15           \\
			& $\eta_2(1645) K$                            &        4.02           \\
			& $\eta_2(1870) K$                            &       0.07           \\
			& $\eta K_2(1820) $                     &        0.05           \\
			& $\eta K_2(1770)$                           &        $<0.01$         \\
			$ 0^+\rightarrow1^-+2^+ $     & $K^*(892)f_2(1270) $                     &       0.93          \\
			& $K^*(892)a_2(1320) $                     &        12.56           \\
			
			& $K_2^*(1430)\rho$                 &       17.92          \\

			& Total width                                 &         180.23   \\           	
			\hline\hline
			
		\end{tabular}
	\end{center}
\end{table}

\section{SUMMARY}
\label{sec:summary}
In this work, we have discussed the possible assignments of $\kzst(1950)$ and $\kzst(2130)$ by calculating the strong decay widths within the $\spl$ strong decay model.

We suggest that the $K_0^*(2130)$ could be assigned as $K_0^*(3^3P_0)$ based on its mass and width.
However, the $K_0^*(1950)$ seems like an exotic state,  because its width can not be reasonably reproduced within the $^3P_0$ model.

With the assignment of the $\kzst(2130)$ as the $\kzst(3\spl)$ state, we have roughly estimated the masses of $\kzst(2\spl)$ and $\kzst(4\spl)$ to be about 1811 MeV and 2404 MeV, respectively, within the Regge trajectories. The total width of $\kzst(2\spl)$ is predicted to be about 656 MeV, which implies that this state is not easy to be observed experimentally.
The total width of $\kzst(4\spl)$ is predicted to be about 180 MeV, which could  be helpful to search for the $\kzst(4\spl)$ state in future.

\section{Acknowledgements}
This work is partly supported by the China Postdoctoral Science Foundation Funded Project under Grand No.2021M701086, and supported by the Natural Science Foundation of Henan under Grand No.212300410123 and No.222300420554. It is also supported by the National Natural Science Foundation of China under Grant No.12192263. It is also supported by the Key Research Projects of Henan Higher Education Institutions under No.20A140027, the Project of Youth Backbone Teachers of Colleges and Universities of Henan Province (2020GGJS017), the Youth Talent Support Project of Henan (2021HYTP002), the Fundamental Research Cultivation Fund for Young Teachers of Zhengzhou University (JC202041042), and the Open Project of Guangxi Key Laboratory of Nuclear Physics and Nuclear Technology, No.NLK2021-08.

\end{document}